\documentclass[aps,prb,twocolumn,showpacs,eqsecnum]{revtex4}
\usepackage{euscript}
\usepackage{amssymb}
\usepackage{amsmath}
\usepackage{graphicx}
\begin{document}

\title{Magnetic field effects on electron-hole recombination in disordered
organic semiconductors}

\author{A. I. Shushin}
\affiliation{Institute of Chemical Physics, Russian Academy of
Sciences, 117977, GSP-1, Kosygin str. 4, Moscow, Russia}

%e-mail: shushin@chph.ras.ru}

\begin{abstract}
Characteristic properties of magnetic field effects on spin
selective geminate and bulk electron-hole polaron pair (PP)
recombination are analyzed in detail within the approach based on
the stochastic Liouville equation. Simple expressions for the
magnetic field (B) dependence of recombination yield and rate are
derived within two models of relative PP motion: free diffusion and
diffusion in the presence of well (cage). The spin evolution of PPs
is described taking in account the relaxation induced by hyperfine
interaction, anisotropic part of the Zeeman interaction induced, as
well as $\Delta g$-mechanism. A large variety of the $B$-dependences
of the recombination yield $Y(B)$ and rate $K(B)$ is obtained
depending on the relative weights of above-mentioned mechanisms. The
proposed general method and derived particular formulas are shown to
be quite useful for the analysis of recent experimental results.
\end{abstract}


\maketitle

\bigskip

\section{Introduction}

Magnetic field effects (MFEs) on various processes in organic
semiconductors are actively studied for many
years.\cite{Mer1,Mer2,Fr1,Prig1,Var0,Var1,Woh1,Prig2,Woh2a,Woh2,
Var2,Prig3,Var3,Woh3,Maj,Var4,Woh4} These studies concern different
types of  MFEs: the magnetic field dependent
photoconductivity,\cite{Fr1} and
photoluminescence,\cite{Var0,Woh2a},
magnetoelectroluminescence,\cite{Var2,Var3}
magnetoresistance,\cite{Prig1,Var1,Woh1,Prig2,Woh2a,Woh2,Var2,
Prig3,Var3,Woh3,Maj,Var4,Woh4} etc.

The mechanisms of MFEs in these processes are the subjects of hot
debates for many years. It is, however, widely
accepted,\cite{Sw,St,Tal} that a large number of MFEs result from
the effect of the magnetic field ($B$) on the spin selective
reactions with participation of paramagnetic particles: polarons
($P$) and triplet excitons ($T$). In a large number of them the key
stage is the recombination of pairs of electron ($e$) and hole ($h$)
polarons, i.e. particles with electronic spin
$1/2$,\cite{Prig1,Prig2,Woh3,Woh4} which are called hereafter
polaron pairs (PP). The $T-P$ quenching $T-T$ annihilation are also
believed be the important spin selective process which can give
significant contribution to the observed MFEs in organic
semiconductors.\cite{Sw,Gi1,Gi2,Gi3}

There are some reviews of experimental and theoretical works on MFEs
in organic semiconductors.\cite{Sw,St,Tal} Recent extensive
experimental investigations, however, inspire further theoretical
studies of MFEs.\cite{Prig1,Prig2,Woh3,Woh4} Despite evident
progress in these studies there are still many problems to be
discussed.

In particular, close attention has been attracted to the $e\!-\!h$
PP recombination mechanism of MFEs in disordered
semiconductors.\cite{Sw,Prig1,Woh4} The important problem consists
in proper treatment of the effect of polaron migration and disorder
of spin dependent interactions, giving rise to the MFEs. The
majority of theoretical works are mainly based either on numerical
or somewhat simplified analytical description of the spin/space
evolution of PPs,\cite{Sw,Woh3,Woh4} though fairly high accuracy of
above mentioned recent measurements motivates more detailed
theoretical investigations, which could allow for obtaining
sufficiently accurate and rigorous formulas for the MFEs.

In this work we discuss the method of describing specific features
of the PP recombination mechanism of MFEs in disordered organic
semiconductors. The method is based on the diffusion approximation
for hopping polaron migration, fairly reasonable at long times of
MFE formation. The hopping motion is assumed to lead not only to
stochastic spatial evolution but also to fluctuations of disordered
spin dependent hyperfine interaction (HFI) and anisotropic part of
the Zeeman interaction (AZI), resulting from the anisotropy of
$g$-factors of polarons. The fluctuating HFI and AZI give rise to
spin relaxation, which is described by Bloch-type equations, valid
in the realistic limit of hopping rates much larger than these
interactions (in frequency units).

The kinetics of MFE generation is determined by the spin/space
evolution of PPs, which is described by the PP spin density matrix.
In the above-formulated approximations this matrix satisfies the
stochastic Liouville equation (SLE).\cite{Fre,St,Shu1}

With the SLE approach we analyze the properties of MFE for geminate
and bulk processes within two models of PP relative motion: free
diffusion and diffusion in the presence of a potential
well.\cite{Shu2,Shu2a} The PP spin evolution is described taking
into consideration the above-mentioned HFI and AZI induced
relaxation, as well $\Delta g$-mechanism.\cite{St} In these two
models of relative motion simple expressions are derived for the
MFE, i.e. for $B$ dependent recombination yield $Y(B)$ and rate
$K(B)$ (in geminate and bulk precesses, respectively).

The analysis with obtained formulas reveals different types of
$Y(B)$ and $K(B)$ behavior, as $B$ increased, for HFI and AZI
induced relaxation mechanisms: decreasing and decreasing,
respectively. The combination of these mechanisms is found to result
in a large variety of non-monotonic $Y(B)$ and $K(B)$ dependences.
The extra contribution of $\Delta g$-mechanism can lead to some
additional specific features of MFEs behavior at large magnetic
fields $B$ whose specific features appear to depend on the mechanism
of relative motion of polarons.

In our discussion we also concern some possible applications of
obtained expressions to the interpretation of some recent
experimental results.

\section{Mechanism of polaron migration}

The characteristic properties of MFEs on the migration assisted PP
recombination are, naturally, essentially determined by the
mechanism of polaron migration.

There are a number of models of migration in disorder
semiconductors. One of the most popular is the Miller-Abrahams
model.\cite{Abr,Bob1} In this model the rate $w_{ij}$ of hopping
from the site $j$ to the site $i$ is written as $w_{i j} = w_{_{E_i
E_j}} = w_0 e^{- \theta(E_i - E_j)(E_i - E_j)/(k_B T)},$ where $w_0$
is the characteristic rate constant, $\theta (x)$ is the Heaviside
step function, $r_{ij}$ is the distance between these sites, and
$E_j$ and $ E_i$ are the energies of initial and final states,
respectively, assumed to be randomly distributed parameters whose
(broad) distribution functions are determined by a number of intra
and interpolaron interaction.\cite{Abr,Bob1}

Even in this relatively simple model the kinetics of the space/time
evolution of charge carriers can be obtained only numerically.

It is worth noting, however, that the MFEs are determined by very
long times of order of the characteristic time $\tau_{_S} \sim
10^{-8} s^{-1} $ of spin evolution of  $e$- and
$h$-polarons,\cite{Shu1} which is much longer than the average
hopping time $w_0^{-1}$: $\tau_{_S} \gg w_0^{-1}$ (see Sec. VIII).
At times $t \gtrsim \tau_{_S} \gg w_0^{-1}$ in the wide region of
parameters of the model the space/time evolution is fairly
reasonably described by the diffusion approximation. One of
indications of this fact is a reasonably good accuracy of the
Langevin formula for the recombination rate.\cite{Bob1}

In the diffusion approximation the migration kinetics of
$\nu$-polaron ($\nu = e, h$) is described by the time dependent
probability distribution function $p_{\nu} ({\bf r}_{\nu},t)$ in the
continuum space $\{ {\bf r}_{\nu}\}$, obeying the Smoluchowski
equation
\begin{equation} \label{mot1a}
\dot p_{\nu} = D_{\nu} \nabla_{{\bf r}_{\nu}} (\nabla_{{\bf
r}_{\nu}} p_{\nu} + p_{\nu} \nabla_{{\bf r}_{\nu}}u_{\nu} ),\;\;
(\nu = e, h),
\end{equation}
In eq.(\ref{mot1a}) $D_{\nu}^{} \sim l_{\nu}^2 w_{0_{\nu}}^{}$ is
the effective diffusion coefficient for $\nu$-polaron, in which
$l_{\nu}^{}$ and $w_{0_{\nu}}$ are the characteristic hopping length
and hopping rate. Note that eq. (\ref{mot1a}) is actually a
selfconsistent equation, which incorporates many particle effects,
showing themselves in possible dependence of the diffusion
coefficient on the concentration of polarons, as well as
concentration and temperature dependence of the effective potential
$u({\bf r}_{\nu})$.\cite{Bob1}

Our further analysis of MFEs will be based on the diffusion
approximation (\ref{mot1a}). Nevertheless, some specific features of
hopping kinetics at short times, predicted by the Miller-Abrahams
model, appear to be important as well for the description of MFEs
(see below).

\section{Spin evolution}

Quantum (spin) PP evolution plays the key role in the generation of
MFEs on PP recombination. PP recombination is known to be a spin
selective process with the rate depending on the total electron spin
of the pair of $e$- and $h$-polarons:  ${\bf S} = {\bf S}_e + {\bf
S}_h$, where ${\bf S}_{\nu}$ is the spin of $\nu$-polaron ($\nu = e,
h$). In our discussion we will assume that the rate is non-zero only
in the singlet ($S$) state corresponding to the total PP spin $S =
0$.\cite{Sw,St}

The spin evolution of $e$- and $h$-polarons is described by spin
density matrices $\rho_{\nu}, \:(\nu = e, h),\,$ in the two state
Hilbert spaces, in which two states $|\nu_{\pm}\rangle$ correspond
to two spin projections onto $z$-axis, taken to be parallel to the
vector ${\bf B}$ of external magnetic field:
$S_{\nu_z}|\nu_{\pm}\rangle = \pm \frac{1}{2} |\nu_{\pm}\rangle$. In
general, these matrices can be represented in terms of expansion in
bilinear combinations of the states
\begin{equation} \label{gen0}
|\nu_{\mu\mu'}\rangle = |\nu_{\mu}\rangle\langle \nu_{\mu'}|.
\end{equation}
It is important to note that $|\nu_{\mu\mu'}\rangle$  can  be
considered as vectors in some space, called hereafter the Liouville
space and denoted as $\{\nu_{\mu\mu'}\!\}$, in which the conjugated
vectors $\langle \nu_{\mu'\mu} |$ are defined by the relation
$\langle \nu_{\mu_1^{}\mu_1'} |\nu_{\mu_2^{}\mu_2'}\rangle =
\delta_{\mu_1^{}\mu_2^{}}\delta_{\mu_1'\mu_2'}$. Noteworthy is also
that this definition implies linear independence of the vectors
$|\nu_{\mu\mu'}\rangle$ and $|\nu_{\mu'\mu}\rangle$, i.e. $\langle
\nu_{\mu'\mu} |\nu_{\mu\mu'}\rangle = 0$. In
$\{\nu_{\mu\mu'}\!\}$-space, the density matrices
%$\rho_{\nu}$
$\rho_{\nu}^{} =  \sum\nolimits_{\!_{\mu,\mu'=
\pm}}\!\rho_{\nu_{\mu\mu'}}^{}|\nu_{\mu}\rangle\langle \nu_{\mu'}| $
can be written as a vectors:
\begin{eqnarray} \label{gen1}
\rho_{\nu}^{}& \equiv & |\rho_{\nu}^{}\rangle =
\sum\nolimits_{\!_{\mu,\mu'=
\pm}}\!\rho_{\nu_{\mu\mu'}}^{}|\nu_{\mu\mu'}\rangle.
\end{eqnarray}
In addition to vectors, in the Liouville space one can also
introduce operators, which will be called superoperators.

The time evolution of the spin density matrix $\rho_{\nu}$ satisfies
the Schr\"odinger equation (hereafter we put $\hbar = 1$, i.e.  use
frequency units for energy parameters)
\begin{equation} \label{gen2}
\dot \rho_{\nu}^{} = -i \hat H_{\nu}^{}\rho_{\nu}. % - \hat
%W_{\!\nu}^{} \rho_{\nu}^{}.
\end{equation}
In this equation $\hat H_{\nu}$ the spin Hamiltonian superoperator
(operator in the Liouville space) expressed in terms of the
Hamiltonian $H_{\nu}$ (in the Hilbert space) by the relation
\begin{equation} \label{gen2a}
\hat H_{\nu} \rho_{\nu} \equiv \hat H_{\nu} |\rho_{\nu}\rangle =
H_{\nu}\rho_{\nu} - \rho_{\nu} H_{\nu}.
\end{equation}
For matrix elements of $\hat H_{\nu}$ we get $\langle
\nu_{\mu_1^{}\mu_1'}| \hat H_{\nu} |\nu_{\mu_2^{}\mu_2'} \rangle =
\langle \nu_{\mu_1^{}}|H_{\nu}|\nu_{\mu_2^{}} \rangle
\delta_{\mu_1'\mu_2'} - \langle \nu_{\mu_2'}|H_{\nu} |\nu_{\mu_1'}
\rangle \delta_{\mu_1^{}\mu_2^{}} $.

\section{Spin Hamiltonian and spin relaxation}

In general the spin Hamiltonian $H_{\nu}({\bf r}_{\nu}^{})$, ($\nu =
e, h$), depends on the coordinate ${\bf r}_{\nu}$ of $\nu$-polaron,
due to inhomogeneity of the medium. This dependence results in
fluctuations of $\hat H_{\nu} (t) \equiv \hat H_{\nu} \big({\bf
r}_{\nu}(t)\big)$ caused by hopping motion of $\nu$-polaron, i.e.
stochastic changing ${\bf r}_{\nu}(t)$. The Hamiltonian can
conventionally be represented as a sum of the stationary and
fluctuating parts:
\begin{equation} \label{ham0}
H_{\nu}^{}({\bf r}_{\nu}^{}(t)) = \bar H_{_{Z_{\nu}}} +
H_{_{H_{\nu}}}({\bf r}_{\nu}^{}(t)) + H_{_{A_{\nu}}}({\bf
r}_{\nu}^{}(t)).
\end{equation}
in which the stationary part
\begin{equation} \label{ham0a}
\bar H_{_{Z_{\nu}}} = \bar g_{\nu}\beta ({\bf S}_{\nu}^{}\cdot{\bf
B}) = \omega_{\nu}^{}S_{\nu_z^{}}^{}
\;\;\mbox{with}\;\;\omega_{\nu}^{} = \bar g_{\nu}\beta B
\end{equation}
is the average Zeeman interaction of the electron spin ${\bf
S}_{\nu}$ with the magnetic field ${\bf B}$. The fluctuating parts
$H_{_{H_{\nu}}}\!({\bf r}_{\nu}^{}(t))$ and $H_{_{A_{\nu}}}\!({\bf
r}_{\nu}^{}(t))$ represent HFI and AZI contributions, respectively.

Fluctuating HFI and AZI induce spin relaxation, which can be treated
within the short correlation time approximation.\cite{Abr1} In this
approximation the relaxation is described by the Bloch-Redfield
equations for spin density matrices of polarons.\cite{Abr1}  The
relaxation kinetics is determined by the relaxation supermatrices of
the form
\begin{equation} \label{ham0b}
\hat W_{Q_\nu}^{} =  w_{_{\!Q_\nu}}^p\hat P_{\nu}^p  +
w_{_{\!Q_\nu}}^{n_{}} \hat P_{\nu}^n, \;\;\;(Q = H, A),
\end{equation}
for both HFI ($H$) and AZI ($A$) mechanisms, where
\begin{eqnarray} \label{ham0c}
\hat P_{\nu}^{p} \,&=& %\mbox{$\frac{1}{2}$}
\big(|\nu_{{_{++}}}\rangle-|\nu_{_{--}}\rangle\big)\big(\langle
\nu_{_{++}}|-\langle \nu_{_{--}}|\big),\\
\hat P_{\nu}^{n} &=&|\nu_{_{+-}}\rangle\langle \nu_{_{+-}}| +
|\nu_{_{-+}}\rangle\langle \nu_{_{-+}}|.
\end{eqnarray}
are the operators in the subspaces of diagonal ($\hat P_{\nu}^{p}$)
and non-diagonal ($\hat P_{\nu}^{n}$) elements of the density
matrix. The first and second terms in eq. (\ref{ham0b}) describe the
population and phase relaxation, respectively, with rates\cite{Car1}
\begin{equation} \label{ham0d}
w_{_{\!Q_\nu}}^p = \bar w_{_{\!Q_\nu}}^{} {\cal J}_{_{Q}}^{\nu}
(\omega_{\nu}), \;\;\; w_{_{\!Q_\nu}}^n = \bar w_{_{\!Q_\nu}}^{}
[p_{_Q}^{} + {\cal J}_{_{Q}}^{\nu} (\omega_{\nu})].
\end{equation}
Here $p_{_Q}^{}$ is the numerical parameter, depending on the
relaxation mechanism (specified by  $Q = H, A$), $\, \bar
w_{_{\!Q_\nu}}^{} $ is the characteristic rate (see below), and
\begin{equation}\label{ham0e}
{\cal J}_{_Q}^{\nu} (\omega) = %\frac{1}{\tau_{_{\!Q_\nu}}}\!
\tau_{_{\!Q_\nu}}^{-1} \int_0^{\infty}\!\!\! dt \, \Phi_{\!_Q}^{\nu}
(t) \cos (\omega t)
\end{equation}
with
\begin{equation}\label{ham0g}
\tau_{_{\!Q_\nu}}^{}  = \int_0^{\infty}\!\! dt \, \Phi_{\!_Q}^{\nu}
(t)\end{equation} is the normalized Fourier transformed correlation
function satisfying the relation ${\cal J}_{\nu} (0) = 1$.

The form of functions  ${\cal J}_{_Q}^{\nu} (\omega)$ depends on the
mechanism of fluctuations. In the considered model of
polaron-hopping induced fluctuations these functions are essentially
determined by hopping kinetics and specific features of the
orientational distribution of molecules responsible for the HFI and
AZI. In our work we will need only the most general properties of
${\cal J}_{_Q}^{\nu} (\omega)$. They can be understood in a simple
model of random Hamiltonians $H_{_{H_{\nu}}}\!({\bf r}_{\nu}^{})$
and $H_{_{A_{\nu}}}\!({\bf r}_{\nu}^{})$ uncorrelated at different
sites (with zero mean values), in which $\Phi_{_Q}^{\nu} (t) \sim
\Phi_{\nu} (t) = \big\langle \exp \big(\!-\! w_{_{E_iE_j}}|t|\big)
\big\rangle_{\!_{E_i E_j}} $ and
\begin{equation} \label{ham0f}
{\cal J}_{_{H}}^{\nu} (\omega) = \big\langle
w_{_{E_iE_j}}\big(w_{_{E_iE_j}}^2 + \omega^2\big)_{}^{-1}
\big\rangle_{\!_{E_i E_j}}/\big\langle
w_{_{E_iE_j}}^{-1}\big\rangle_{\!_{E_i E_j}}.
\end{equation}

\subsection{Relaxation mechanisms}

\subsubsection{HFI and HFI induced relaxation}

The HFI $H_{_{H_{\nu}}}$ is determined by the spin-spin interaction
of the electron and paramagnetic nuclei  $\nu_j$ (with spin
$I_{\nu_{j}}$), localized in close surrounding of electronic spins
${\bf S}_e$ and ${\bf S}_h$ of $e$- and $h$-polarons. In the
realistic case of a large number of nuclei the interaction can quite
accurately be approximated by that of spins ${\bf S}_e$ and ${\bf
S}_h$ with (classical) random magnetic fields ${\bf B}_e$ and ${\bf
B}_h$, respectively, whose distributions are isotropic and Gaussian
with mean squares $\langle {B}_{\nu}^2 \rangle \sim \sum_{j}
I_{\nu_{j}}(1+I_{\nu_{j}}) a_{\nu_{j}}^2,\: (\nu = e, h)$,
determined by hyperfine coupling constants $a_{\nu_{j}} $.\cite{Wol}

Hopping of $e$- and $h$-polarons results in sudden changing of
nuclear magnetic field, which can be considered as a stochastic
vector ${\bf B}_{\nu}  (t), \; (\nu = e, h),\:$ with $\langle {\bf
B}_{\nu} \rangle = 0$ and the correlation function of projections
${B}_{\nu_{q}} (t), \: (q = x,y,z)$: $\langle {B}_{\nu_{q}}(t)
{B}_{\nu_{q'}}(0)\rangle = \mbox{$\frac{1}{3}$} \delta_{qq'}\langle
{B}_{\nu}^{2} \rangle \Phi_{\!_H}^{\nu} (t)$.

The HFI mechanism predicts the relaxation superoperator $\hat
W_{H_\nu}^{}$ of the form (\ref{ham0b}) with the rates
$w_{_{\!H_\nu}}^p $ and $w_{_{\!H_\nu}}^n$ given by eq.
(\ref{ham0d}), in which
\begin{equation} \label{ham7}
p_{_Q} = 1 \;\;\; \mbox{and} \;\;\; \bar w_{_{\!H_\nu}}^{} =
\mbox{$\frac{1}{3}$}(g_{\nu}\beta)^2\langle {B}_{\nu}^2 \rangle
\tau_{_{H_{\nu}}},
\end{equation}
with the correlation time $\tau_{_{H_{\nu}}}$ defined in eq.
(\ref{ham0g}).

\subsubsection{AZI and AZI induced relaxation}

The AZI $H_{A}^{}$ results from the deviation of $g_{\nu}$-factors
of $e$- and $h$-polaron spins from the free electron value $g_0 =
2$.\cite{Abr1,Car1} In general, $g$-factors are actually $\hat
g_{\nu}$-tensors\cite{Car1} whose eigenvectors are determined by the
geometry of molecules, at which $e$- and $h$-polarons are located.
The $\hat g_{\nu}$-tensors are conveniently represented as sums of
isotropic ($\bar g_{\nu}$) and anisotropic ($\delta \hat g_{\nu}$)
parts:
\begin{equation} \label{ham9}
\hat g_{\nu} = \bar g_{\nu} + \hat g'_{\nu}, \;\; \mbox{where} \;\;
\bar g_{\nu}  = {\rm Tr} (\hat g_{\nu}).
\end{equation}
The representation (\ref{ham9}) implies that ${\rm Tr} (\hat
g'_{\nu}) = 0$. Usually, for systems under study eigenvalues of
$\hat g_{\nu}$-tensors are close to $g_0 = 2$: $\delta \bar g_{\nu}
= \bar g_{\nu} - g_0 \sim \| \hat g'_{\nu}\| \lesssim 10^{-2}$.

The AZI part of the interaction of the electron spin $\nu$ ($\nu =
e, h$) can be defined as
\begin{equation} \label{ham10}
H_{A_{\nu}}^{} = \beta {\bf S}_{\nu} \hat g_{\nu} {\bf B} - \bar
H_{Z_{\nu}}^{} = \beta {\bf S}_{\nu} \hat g'_{\nu} {\bf B}.
\end{equation}

In the AZI induced relaxation mechanism the rate superoperator $\hat
W_{A_\nu}^{}$ is written as (\ref{ham0b}) with rates
$w_{_{A_\nu}}^p$ and $w_{_{A_\nu}}^n$ (\ref{ham0d}), in
which\cite{Car1}
\begin{equation} \label{ham12}
p_{_Q} = \mbox{$\frac{4}{3}$} \;\;\; \mbox{and} \;\;\; \bar
w_{_{A_\nu}}^{} = \mbox{$\frac{1}{10}$}(\hat g'_{\nu}\!\!: \hat
g'_{\nu}) (\beta {B})^2 \tau_{_{A_\nu}}^{},
\end{equation}
with $\tau_{_{A_{\nu}}}$ given by eq. (\ref{ham0g}).

\subsection{Hamiltonian and relaxation in polaron pairs}

The PP spin Hamiltonian is, in general, a complicated function of
coordinates ${\bf r}_h$ and ${\bf r}_e$, determined by the electron
spin dependent PP interactions: spin exchange and dipole-dipole
interactions. Fortunately, the their effect on majority of MFEs is
weak and can be neglected.\cite{St} In such a case the PP spin
Hamiltonian can be represented as a sum of spin Hamiltonians of $e$-
and $h$-polarons:
\begin{equation}\label{ham17}
H = \bar H_{_{Z_e}} + \bar H_{_{Z_h}} = \bar g \beta B {S}_{z}^{} +
\mbox{$\frac{1}{2}$} \Delta \bar g \beta B ({S}_{e_z^{}}^{}\!\! - {
S}_{h_{z}^{}}^{}),
\end{equation}
where ${S}_z^{} = {S}_{e_z^{}}^{}\! + { S}_{h_{z}^{}}^{}$,
\begin{equation}\label{ham17a}
\bar g = \mbox{$\frac{1}{2}$} (\bar g_e^{} + \bar g_h^{})
\;\;\mbox{and}\;\; \Delta\bar g = \bar g_e^{} - \bar g_h^{}.
\end{equation}

Within similar assumptions, the PP spin relaxation can be described
by the superoperator, which is also a sum of corresponding
operators for separate polarons: %$\nu = e, h$ [see Sec. IV.A]
\begin{equation} \label{ham19}
\hat W = \hat W_e + \hat W_h = \sum\nolimits_{\nu = e,h} (\hat
W_{_{H_\nu}}^{} + \hat W_{_{A_\nu}}^{}).
\end{equation}

The PP spin evolution is described in the four-state Hilbert space.
In principle, one can use any basis in this space, for example the
basis of states $|e_{\mu}\rangle |h_{\mu'}\rangle,\:(\mu,\mu' =
\pm),$ of the pair of non-interacting polarons. However, in what
follows it will be more convenient to use the basis of eigenstates
of the $z$-projection, $S_z^{}$, of the vector ${\bf S}$:
$|S\rangle, |T_{\mu}^{}\rangle, \:(\mu = 0,\pm)$, which correspond
to the total; spin $S = 0$ ($S$) and $S = 1$ ($T$):
\begin{eqnarray}
|S\rangle\,\,\, &=& \mbox{$\frac{1}{\sqrt{2}}$}(|e_{+}\rangle
|h_{-}\rangle - |e_{-}\rangle |h_{+}\rangle),\label{ham19a}\\
|T_0^{}\rangle\, &=& \mbox{$\frac{1}{\sqrt{2}}$}(|e_{+}^{}\rangle
|h_{-}^{}\rangle + |e_{-}^{}\rangle |h_{+}^{}\rangle),\label{ham19b}\\
|T_{\pm}^{}\rangle &=& |e_{\pm}\rangle |h_{\pm}\rangle.
\label{ham19c}
\end{eqnarray}
In addition to the states in the Hilbert space it is worth
introducing the states in the Liouville space
\begin{equation} \label{ham20}
| XY \rangle = | X \rangle \langle Y|,\;\; (X, Y = S, T_{\mu}^{}).
\end{equation}

\section{Stochastic Liouville equation}

The specific features of PP recombination processes (both geminate
and bulk) can, in general, be expressed in terms the PP density
matrix  $\rho ({\bf r},t)$. In the case of Markovian $e\!-\!h$
relative motion  $\rho ({\bf r},t)$ is known to satisfy the
SLE,\cite{Fre,Shu1} which within the diffusion approximation is
written as
\begin{equation}\label{mfe1}
\dot \rho = -[\hat \Lambda + \hat K(r)  + \hat {L}]\rho,
\;\;\mbox{where}\;\; \hat \Lambda =  i\hat H + \hat W,
\end{equation}
$\hat H  = [H, \dots]$ is the superoperator representation of the PP
spin Hamiltonian $ H$ [see eq. (\ref{ham17})] in the Liouville
space, and $\hat W$ is the superoperator of PP spin relaxation,
defined in eq. (\ref{ham19}).

The relative diffusive motion of $e$- and $h$-polarons is described
by the Smoluchowski operator $\hat L$, defined by the expression
\begin{equation}\label{mfe5}
\hat {L}\rho = -D \nabla_{\bf r} (\nabla_{\bf r}\rho + \rho
\nabla_{\bf r} u ),
\end{equation}
in which ${\bf r} = {\bf r}_h^{}-{\bf r}_e^{}$ the relative PPq
coordinate,  $D = D_h + D_e$ is the relative diffusion coefficient
[see eq. (\ref{mot1a})],  and $u (r) = U(r)/(k_B T)$ is the
dimensionless PP interaction potential assumed to depend only on the
distance $r = |{\bf r}|$.

In our analysis we will consider the spherically symmetric problem,
thus reducing it to studying the evolution along the radial
coordinate $r = |{\bf r}| = |{\bf r}_h^{}-{\bf r}_e^{}|$ only.

The potential $u (r)$ is assumed to be of the shape of potential
well (see Fig. 1). This potential is characterized by the barrier,
of height $u_r^{}= u(d) - u(r_b) > 1 $, at the distance $d$ of
closest approach of polarons (which models the suggested smallness
of the reaction rate), the coordinate $r_b$ of the bottom ($r_b >
d$), the well depth $u_b^{} = -u(r_b) > 1$, as well as the Onsager
radius
\begin{equation} \label{mfe6}
l_c = \int_{r_b}^{\infty}\!\!dr\,r^{-2}e^{u(r)}.
\end{equation}

\begin{figure}
\setlength{\unitlength}{1cm}
\includegraphics[height=3.7cm,width=7cm]{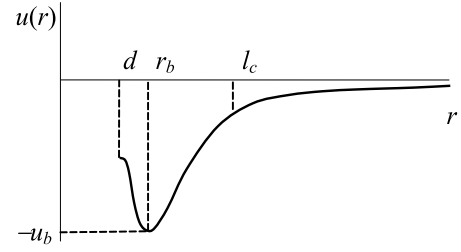}% Here is how to import EPS art
\caption{Schematic picture of the PP interaction potential $u (r) =
U(r)/(k_B^{} T)$, in which $d$ is the distance of closest approach,
$r_b$ is the coordinate of the well bottom, $u_b^{} = -u(r_b)$ is
the well depth, and $l_c^{}$ is the Onsager radius.}
\end{figure}

The term $\hat K(r)\rho $ describes spin selective recombination
assumed to occur only in $S$-state of the PP:\cite{St,Fre}
\begin{equation}
\hat K (r)|\rho\rangle = \mbox{$\frac{1}{2}$}\kappa (r) (
{P}_S^{}\rho + \rho{P}_S^{})\label{mfe7a} \equiv \kappa (r) \hat
{\cal P}_{\!_S}|\rho\rangle,
\end{equation}
where $\kappa (r)$ is the distance dependent recombination rate. In
eq. (\ref{mfe7a}) ${ P}_S = |S\rangle \langle S|$ is the operator of
projection on the $S$-state and
\begin{equation}\label{mfe7b}
\hat {\cal P}_{\!_S} = \hat P_{SS}^{} + \mbox{$\frac{1}{2}$}
\sum\nolimits_{\mu=0,\pm}\!(\hat P_{ST_{\mu}}^{} + \hat
P_{T_{\mu}S}^{})
\end{equation}
is the superoperator, controlling the spin dependence of reactivity
(and accompanying dephasing), in which
\begin{equation}\label{mfe7c}
\hat P_{XY}^{} = | XY \rangle\langle XY|,\;\; (X, Y = S,
T_{\mu}^{}),
\end{equation}
are superoperators of projection onto the states (\ref{ham20}) in
the Liouville space. The dependence $\kappa (r)$ is typically short
range and does not affect MFEs very much, so that, for simplicity,
one can apply the contact reactivity model
\begin{equation}\label{mfe8}
\kappa(r) = \kappa_{_S}^{} \delta (r-d), \;\;\mbox{i.e.}\;\; \hat K
(r) = \hat K_{_S}^{} \delta (r-d)
\end{equation}
with $\hat K_{_S}^{} = \kappa_{_S}^{} \hat {\cal P}_{\!_S}^{}$.

The polarons are assumed to reflect at the distance of closest
approach $d$. The process is described by the reflective inner
boundary condition  $(\nabla_r \rho + \rho \nabla_r u)_{r = d}^{} =
0$. In the model of contact reactivity (\ref{mfe8}), however, one
can formally %omit the term $\hat K (r)$ in the SLE (\ref{mfe5}) and
reduce the effect of reactivity to the modification of the inner
boundary condition:
\begin{equation}\label{mfe9}
\big[\nabla_r\rho + \rho \nabla_r u   - (\hat K_{_S}^{}/D)
\rho)\big]_{{r=d}} = 0.
\end{equation}

As for the outer boundary condition it is different for geminate and
bulk processes (see below).

\section{Geminate PP recombination}

The kinetics of geminate PP recombination will be analyzed assuming
that PPs are created with isotropic distribution, localized at a
distance $r = r_i$, and in the spin state, determined by the density
matrix $\rho_i (r)$. This means that the SLE (\ref{mfe1}) should be
solved with the initial condition
\begin{equation}\label{mfe10}
\rho_i^{} (r) = \rho ({\bf r}, t=0) = (4\pi r_i^2)^{-1}\rho_g^{}
\delta (r-r_i^{}).
\end{equation}

The outer boundary condition at $r \to \infty$, is written as
$\rho(r\to\infty,t) \to 0$.

Spherically symmetry of PP interactions, as well as spherical
symmetry of initial and boundary conditions ensures that the spin
density matrix is also spherically symmetric: $\rho({\bf r},t)
\equiv \rho(r,t)$. In this case the PP recombination yield $Y_r$,
can be expressed by formula
\begin{eqnarray}\label{mfe11}
Y_r &=& \int_0^{\infty}\!\!dt \int \!d^3 r\,{\rm Tr} [\hat K(r)\rho
(r,t)]\nonumber\\
&=& 4\pi \kappa_s d^2 \langle S | \widetilde{\rho}_0^{} (d)|S\rangle
\equiv 4\pi \kappa_s d^2 \langle SS | \widetilde{\rho}_0^{}
(d)\rangle,\qquad
\end{eqnarray}
where the matrix $\widetilde{\rho}_0^{} (r)$ is the Laplace
transform $\widetilde{\rho}_{\epsilon} (r) = \!\int_0^{\infty}\!\!dt
\, e^{-\epsilon t}\rho (r,t)$, evaluated at $\epsilon = 0$:
$\widetilde{\rho}_0^{} (r) = \widetilde{\rho}_{\epsilon = 0}^{}
(r)$.

The matrix $\widetilde{\rho}_{0}^{} (r)$ satisfies the steady state
variant of the SLE (\ref{mfe1}
\begin{equation}\label{mfe14}
(\hat \Lambda + \hat {L}_r) \widetilde{\rho}_0^{} = \rho_i (r),
\;\;\mbox{where} \;\; \hat \Lambda = i\hat H + \hat W,
\end{equation}
and $L_r$ is the radial part of the operator $\hat L$ defined by
\begin{equation}\label{mfe15a}
L_r \rho= Dr^{-2}\nabla_r[r^2(\nabla_r \rho + \rho \nabla_r u)].
\end{equation}

Noteworthy is that in the SLE (\ref{mfe14}) we have omitted the
reactivity operator $\hat K(r)$ (\ref{mfe8}), expressing its effect
by the properly chosen inner boundary condition (\ref{mfe9}).

For the initial condition (\ref{mfe10}) the solution of this
equation is easily expressed in terms of the Green's function of the
equation (\ref{mfe14})
\begin{equation}\label{mfe15}
\hat G_{0}^{} = (\hat \Lambda + \hat {L}_r)^{-1}:% \;\;\mbox{with}
\end{equation}
\begin{equation}\label{mfe16}
Y_r = \kappa_s (d/r_i)^2\langle SS |\hat
G_{0}^{}(d,r_i)|\rho_g\rangle.
\end{equation}

In general, $\hat G_{0}^{}(r,r_i)$ can hardly be obtained
analytically. However, fairly simple analytical expressions for
$\hat G_{0}^{}(r,r_i)$ (and thus for $Y_r$) can be found in some
important particular cases.

\subsection{Freely diffusing polarons.}

\subsubsection{General formulas}

In the absence of PP interaction potential [$ u(r) = 0$] the
solution of eq. (\ref{mfe15}) can be obtained in a matrix
form.\cite{Shu1,Shu3} For simplicity we will assume that PPs are
created at a distance of closest approach, i.e. $r_i = d$.

The analytical expression can conveniently be obtained with the use
of the representation
\begin{equation}\label{mfe17}
\hat G_{0}^{} = D^{-1}(r_i/r) \hat g_{0}^{} %\;\;\mbox{with} \;\;
%\hat g_{0}^{} = (\hat k_{0}^2 - \nabla_r^2)^{-1}.
\end{equation}
In this relation
\begin{equation}\label{mfe18}
\hat g_{0}^{} = (\hat k_{0}^2 - \nabla_r^2)^{-1}, \;\;\mbox{with} \;\;
 \hat k_{0} = \big(\hat \Lambda /D\big)^{1/2},
\end{equation}
is the one dimensional Green's function, which satisfies the inner
boundary condition similar to (\ref{mfe9}):
\begin{equation}\label{mfe19}
(\nabla_r - \hat q_s)\hat g_{0}^{}|_{{r=d}} =
0\;\:\mbox{with}\;\:\hat q_s = d^{-1} + (\hat K_{_S}^{}/D).
\end{equation}

The function $\hat g_{0}^{}$ is given by formula\cite{Shu1,Shu3}
\begin{equation}\label{mfe20}
\hat g_{0}^{}(r,r') = \big[e^{-\hat k_{0}^{}(r-d)}\hat \lambda +
e^{-\hat k_{0}^{}|r-r'|}\big](2\hat k_{0}^{})^{-1},
\end{equation}
where
\begin{equation}\label{mfe21}
\hat \lambda = (\hat \theta + 1)^{-1}(\hat \theta -1)e^{-\hat
k_{0}^{}(r'-d)}\;\:\mbox{and} \;\:\hat \theta = \hat q_s^{-1}\hat
k_{0}^{}.
\end{equation}

Substitution of this formula into eqs. (\ref{mfe18}) and
(\ref{mfe11}) results in the expression for the recombination yield:
\begin{equation}\label{mfe22}
Y = Y_{\!{f}}^{} = {\rm Tr}[P_S^{} (\hat {\,\mathbf{P}}_{\!f}^{}
\rho_g^{})] \equiv \langle SS|\hat {\,\mathbf{P}}_{\!f}^{}
|\rho_g^{}\rangle,
\end{equation}
in which
\begin{equation}\label{mfe23}
\hat {\mathbf{P}}_{\!f}^{} = \hat {\cal P}_{r_{\!f}}(1 + d\hat
k_{0}^{} \hat {\cal P}_{e_{\!f}})^{\!-1},
%\;\mbox{with} \; \hat {\cal P}_r = \hat E -
%\hat {\cal P}_e = \hat l/d,
\end{equation}
is the spin dependent supermatrix of reaction/relaxation
probabilities. In eq. (\ref{mfe23})
\begin{equation}\label{mfe23a}
\hat {\cal P}_{r_{\!f}^{}} =  \hat l/d \;\;\mbox{and}\;\; \hat {\cal
P}_{e_{\!f}^{}} = 1 -  \hat l/d,
\end{equation}
are the supermatrices of probabilities of reaction and escaping
(respectively) of PPs, created at $r_i = d$, in the absence of PP
spin evolution. The probabilities are essentially determined by the
supermatrix $\hat l$ of reaction/relaxation radii, corresponding to
the rate supermatrix $\hat K (r)$ [given by eq. (\ref{mfe8})]:
\begin{equation}\label{mfe24}
\hat l = d - \hat q_s^{-1} = d \hat \gamma_{f}^{}/(1\!+\!\hat
\gamma_{f}^{})\;\:\mbox{with}\;\: \hat \gamma_{f}^{} =d\hat
K_{_S}^{}/D.
\end{equation}

Formulas (\ref{mfe22})-(\ref{mfe24}) for the yield of recombination
of freely diffusing polarons are seen to reduce the problem of
evaluation of MFEs to simple matrix operations.

\subsubsection{Fast P-diffusion limit}

The limit of fast relative diffusion (or slow spin evolution), when
$d\|\hat k_{0} \| \ll 1$,\cite{Shu1,Shu3} is  of special interest
for our further analysis. In this limit $\hat
{\mathbf{\,P}}_{\!f}^{} $ can be found in the approximation linear
in $d \hat k_{0}$: $\hat {\mathbf{\,P}}_{\!f}^{} \approx \hat {\cal
P}_{r_{\!f}} (1 - d\hat k_{0}\hat {\cal P}_{e_{\!f}})$, i.e.
\begin{equation}\label{mfe25}
Y_{\!f} %(\rho_g^{})
\approx (l_{SS}/d)p_{_S} - l_{SS}\langle SS|\hat k_{0}\hat {\cal
P}_{e_{\!f}}|\rho_g\rangle,
\end{equation}
where $l_{SS}$ is the reaction radius in $S$-state and $p_{_S} =
\langle SS|\rho_g\rangle $ is the initial population of this state.

For the model (\ref{ham17}), (\ref{ham19}) the yield $Y_{\!f}$ can
be obtained for any initial state $|\rho_g \rangle$ and, in
particular, for
\begin{equation}\label{mfe26}
|\rho_g \rangle  = p_{_S} |\rho_{_S}\rangle + p_{_T}
|\rho_{_T}\rangle,
\end{equation}
in which $p_{_S}$ and $p_{_T}$ are the probabilities of population
of $S$ and $T$ states, respectively ($p_{_S} + p_{_T} = 1$),
\begin{equation}\label{mfe27}
|\rho_{_S}\rangle = |SS\rangle \;\;\mbox{and} \;\; |\rho_{_T}\rangle
= \mbox{$\frac{1}{3}$}\sum\nolimits_{{\mu}} \!
|T_{\mu}T_{\mu}\rangle
\end{equation}

Calculation of $Y_{\!f}$ reduces to evaluating the supermatrix $\hat
k_0 = \big(2\sqrt{\pi D}\,\big)^{-1}\int_0^{\infty}\! dt \,
t^{-3/2}\big(1-e^{-i\hat \Lambda t}\big)$. Substituting thus
obtained  $\hat k_0$ into eq. (\ref{mfe25}), one gets
\begin{equation}\label{mfe28}
Y_{f}^{} \approx {\cal P}_{r_{\!f}^{}}^{_S} \big[p_{_S} +
\mbox{$\frac{1}{12}$}(d k_{_w})\big(p_{_T}{\cal P}_{e_{\!f}^{}}^{_T}
- 3p_{_S}{\cal P}_{e_{\!f}^{}}^{_S}\big)\big],
\end{equation}
where ${\cal P}_{e_{\!f}}^{_T} = 1 - l_{TT}^{}/d = 1, \:{\cal
P}_{r_{\!f}}^{_S} = 1 - {\cal P}_{e_{\!f}}^{_S} = l_{SS}^{}/d, \;$
and
\begin{equation}\label{mfe29}
k_{_w} = \big[ \sqrt{w_{p}^{}}+2{\rm Re}\big(\sqrt{w_{n}^{} + i
\Delta \omega}\,\,\big)\big]/\sqrt{D}.
\end{equation}
In this formula
\begin{equation}\label{mfe30}
\Delta\omega = (\bar g_e - \bar g_h)\beta B = \delta \bar g \beta B.
\end{equation}
is the difference of Zeeman polaron frequencies and
\begin{equation}\label{mfe31}
w_p = 2(w_{_H}^{p} + w_{_A}^{p}) \;\;\mbox{and} \;\;w_n = w_{_H}^{n}
+ w_{_A}^{n},
\end{equation}
are the total rates of population ($w_p$) and phase ($w_n$)
relaxation in the PP, which are the sums of contributions of the HFI
and AZI induced relaxation rates. These contributions are, in turn,
the sums of corresponding relaxation rates in $e$- and $h$-polarons:
\begin{equation}\label{mfe32}
w_{_H}^{q} = w_{_{H_e}}^{q}+ w_{_{H_h}}^{q}; \;\;w_{_S}^{q} =
w_{_{A_e}}^{q} + w_{_{A_h}}^{q}, \; (q = p,n).
\end{equation}
The total rates $w_p$ and $w_n$ are represented as sums of HFI and
AZI induced relaxation rates to clearly reveal two contributions
with different $B$-dependence: decreasing [$w_{_{H_{\nu}}}^{q}
\!(B)$] and increasing [$w_{_{A_{\nu}}}^{q} \!(B)$] as $B$
increases.

\subsection{PP recombination in the presence of interaction}

The PP interaction potential $u (r)$ can essentially affect the MFEs
on PP recombination. For pure repulsive interaction [$u (r)
> 0$] no significant effect is expected except some change of
reaction and relaxation radii, i.e. the elements of the supermatrix
$\hat l$ introduced in eq. (\ref{mfe24}) for freely diffusing
polarons. Much stronger effect is predicted in the case of
attractive interaction, especially for potentials of the shape of
potential well (Fig. 1).\cite{Shu4} In what follows we will discuss
the MFEs in the special case of deep potential well with $u_b \gg
1$.

The analysis can be made with the use of the recently proposed
method of rigorous analytical analysis of the kinetics of diffusion
in the presence of the well.\cite{Shu2} One of important results of
this analysis consists in the fact that in the limit of deep well
(corresponding criterion is given below) the exact SLE (\ref{mfe1})
is equivalent to the model of two kinetically coupled states: the
state within the well (the diffusive cage state), at $d < r < l_c$,
and the state of free diffusion outside the well, i.e. at $r >
l_c$.\cite{Shu2,Shu2a} The spin/space evolution in these two states
are described by density matrices
\begin{equation} \label{mfe35}
n(t) = 4\pi\! \int_d^{l_c} \!\! dr \,r^2 \!\rho (r,t)
\;\:\mbox{and}\;\: \sigma(r,t) = r\rho(r,t),
\end{equation}
respectively, satisfying equations\cite{Shu2,Shu2a}
%\begin{subequations} \label{mfe36}
\begin{eqnarray} \label{mfe36}
\dot n &=&[S_l^{-1}K_{+} \sigma (l_c,t) - (K_- + \hat W_r)n], \qquad
 \label{mfe36a}\\
\dot \sigma &=& [D \nabla_r^2\sigma + (S_lK_{-} n - K_{+}\sigma)
\delta (r-l_c)],\qquad \label{mfe36b}
\end{eqnarray}
%\end{subequations}
in which $S_l = (4\pi l_c)^{-1}$. The terms proportional to
$K_{\pm}$ represent the kinetic coupling (transitions) between the
two states, with transition rates $K_{\pm}$ satisfying the
relations:\cite{Shu2a} $K_{\pm} \rightarrow \infty \quad \mbox{and}
\quad K_{+}/K_{-} = K_e = Z_w,$ where
\begin{equation}\label{mfe38}
Z_w^{} = \int_{d<r<l_c}dr \, r^2 e^{-u(r)}
\end{equation}
is the partition function for the well.

Equations (\ref{mfe36a}) and (\ref{mfe36a}) will be solved assuming
that the PP is initially created within the well, i.e.
\begin{equation} \label{mfe39}
n(0) = \rho_g \quad \mbox{and} \quad \sigma(r,0) = 0.
\end{equation}
As to the boundary conditions for $c(r,t)$, they are given by $l_c
\nabla_r \sigma (r,t) - \sigma(r,t)|_{r=l_c} = 0$ and $\sigma(r
\rightarrow \infty) = 0$.

The term $\hat W_r \widetilde{n}$ in eq. (\ref{mfe36a}) describes
the first order reaction in the well with the supermatrix of rates
\begin{equation} \label{mfe40}
\hat W_{\!r}^{} = \Big(\frac{D}{\kappa_{r}^{}Z_w}\Big) \frac{\hat
\gamma_{c}^{}}{1 + \hat \gamma_{c}^{}} \;\; \mbox{with} \;\;
\kappa_{r}^{} = \!\int_{d}^{r_b} \! \frac{dr}{r^2}\, e^{u(r)}\,
\end{equation}
and $\hat \gamma_{c}^{} = (d\hat K_{_S}^{}/D)(d\kappa_{r}^{}
e^{-u(d)})$.

Solution of eqs. (\ref{mfe36a}) and (\ref{mfe36b}) by the Laplace
transformation in time leads to formula for the recombination
yield\cite{Shu4} similar to eq. (\ref{mfe22}):
\begin{equation}\label{mfe41a}
Y_{c}^{} = {\rm Tr}[P_S^{} (\hat {\,\mathbf{P}}_{\!c}^{\,_{\,}}
\rho_g^{})] \equiv \langle SS|\hat {\,\mathbf{P}}_{\!c}^{}
|\rho_g^{}\rangle,
\end{equation}
but with the supermatrix $\hat {\mathbf{P}}_{\!f}^{}$ replaced by
\begin{eqnarray} \label{mfe41}
\hat {\mathbf{P}}_{\!c}^{} &=& \hat W_r \big[\hat W_c +
\hat \Lambda + W_e (l_c \hat k_{0})\big]^{-1}, \label{mfe41aa}\\
&=&  \hat {\cal P}_{r_{c}}[1 + (\hat \Lambda/W_e  + l_c\hat
k_{0}^{}) \hat {\cal P}_{e_{c}}]^{\!-1}\!,\label{mfe41bb}
\end{eqnarray}
in which $ \: \hat \Lambda = i\hat H + \hat W $, $\, \hat k_0 =
(\hat \Lambda/D)^{1/2},\,$ and
\begin{equation} \label{mfe42}
\hat W_c = \hat W_r + W_e, \;\;\; \mbox{with} \;\; W_e =   Dl_c/Z_w,
\end{equation}
is the supermatrix of total cage decay rates, represented as a sum
of the rate $\hat W_r$ of reaction in the well (cage) and the rate
$W_e$ of escaping from the well.\cite{Shu2,Shu2a} These rates
determine the probabilities of reaction in the well ($\hat {\cal
P}_{r_{c}} $) and escape from the well ($\hat {\cal P}_{e_{c}} $):
\begin{equation}\label{mfe42a}
\hat {\cal P}_{r_{c}} = \hat W_r/\hat W_c \;\;\mbox{and}\;\;\hat
{\cal P}_{e_{c}} = W_e/\hat W_c.
\end{equation}

The detailed analysis shows\cite{Shu2} that expressions
(\ref{mfe41aa}) and (\ref{mfe41bb}) are valid in the limit of not
very fast PP spin evolution, or not very large size of the well
$\delta_c = l_c - d$, when $\delta_c \|\hat k_{0}\| \ll 1$  (this
inequality does not, in principle, mean that $l_c \|\hat k_{0}\| \ll
1$).

According to obtained formulas, the effect of the well manifests
itself in the formation of the diffusive cage, whose evolution is
described as the first order reaction (with the rate $\hat W_r$) and
escaping (with the rate $ W_e$). This simple first order kinetics
is, however, perturbed by the contribution of particles escaped but
recaptured back into the well, resulting in the term $\sim (l_c \hat
k_{0})$ in eqs. (\ref{mfe41aa}) and (\ref{mfe41bb}), which gives
negligibly small contribution to MFEs for deep wells (when $W_e$ is
so small that $\|\hat \Lambda\|/W_e > l_c \|\hat k_{0}\|$), but
strongly affects MFEs in the opposite limit of large escaping rate
$W_e$. Moreover in the limit of fast cage decay (or slow PP spin
evolution), $\|\hat \Lambda\|/W_e \ll l_c \|\hat k_{0}\|$, the
obtained formula reduces to that (\ref{mfe23}) for free
diffusion.\cite{Shu2}

\section{Bulk PP recombination}

The MFEs in bulk recombination are also described by the SLE
(\ref{mfe1}) but with the boundary condition at $r \to \infty$
\begin{equation} \label{bulk1}
\rho (r \to \infty, t) = \rho_{_E} = \mbox{$\frac{1}{4}$}
\sum\nolimits_{{\mu = S,T_{0,\pm}}} \!|\mu \rangle \langle \mu|,
\end{equation}
corresponding to the homogeneous spatial distribution of and the
equilibrium spin states of polarons [represented by the unity matrix
$E_{\!_H} = \sum_{{\mu = S,T_{0,\pm}}} \!|\mu \rangle \langle
\mu|$].

In the case of bulk recombination the observable under study is
$B$-dependent recombination rate $K (t)$. For simplicity we will
discuss the static value $K = K (t\to\infty)$ which can be
represented in the form very similar to that of the expression
(\ref{mfe11}) for the recombination yield:
\begin{eqnarray}\label{bulk2}
K &=& \int \!d^3 r\,{\rm Tr} [\hat K(r)\rho
(r,t\to\infty)]\nonumber\\
&=& 4\pi \kappa_s d^2 \langle S | \widetilde{\rho}_0^{} (d)|S\rangle
\equiv 4\pi \kappa_s d^2 \langle SS | \widetilde{\rho}_0^{}
(d)\rangle,\qquad
\end{eqnarray}
This formula is written, taking into account that the equation for
the static solution $\rho_{st} (r)$ coincides with the Laplace
transform $\widetilde{\rho}_{\epsilon =0}^{} (r) \equiv
\widetilde{\rho}_0^{} (r)$ [however, with the outer boundary
condition (\ref{bulk1})]: $\rho_{st} (r) = \widetilde{\rho}_0^{}
(r)$.

\subsection{Freely diffusing polarons.}

The stationary solution $\widetilde{\rho}_0^{} (r)$ of the SLE
(\ref{mfe1}) satisfying eq. (\ref{mfe14}) with $\rho_i= 0$ can be
found by solving more simple equation for $\widetilde{\sigma}_0^{}
(r) = r^{-1}\widetilde{\rho}_0^{} (r)$
\begin{equation} \label{bulk3}
(\hat k_{0}^2 - \nabla_r^2)\widetilde{\sigma}_0^{} = 0
\;\;\mbox{with}\;\;\hat k_{0} = \big(\hat \Lambda/D\big)^{1/2}.
\end{equation}
The function $\widetilde{\sigma}_0^{}(r)$ satisfies the inner
boundary condition $(\nabla_r - \hat
q_s)\widetilde{\sigma}_0^{}|_{{r=d}} = 0$, similar to (\ref{mfe19}),
and the outer one, which, according to eq. (\ref{bulk1}), is written
as
\begin{equation} \label{bulk4}
\widetilde{\sigma}_0^{}(r \to \infty) = r \rho_{_E}.
\end{equation}

By direct substitution into eq. (\ref{bulk3}) one can show that the
solution of this equation is given by
\begin{equation} \label{bulk5}
\widetilde{\sigma}_0^{}(r) = \big[-e^{-\hat k (r-d)} (1 + \hat
\theta)^{-1}\hat l + r \big]\rho_{_E} .
\end{equation}
where $\hat \theta$ is defined in eq. (\ref{mfe21}). With the use of
this solution the recombination rate is represented as
\begin{equation} \label{bulk6}
K = K_{{f}}^{} = K_{f}^{0} Y_{f_E^{}}^{}, \;\,\mbox{where}\;\,
%{\rm Tr}[\hat P_{S} (\hat {\,\mathbf{P}}_{\!f}^{} \rho_{_E})]
Y_{f_E^{}}^{} = \langle SS| \hat {\,\mathbf{P}}_{\!f}^{}
|\rho_{_E}\rangle,
\end{equation}
In this formula $Y_{f_E^{}}^{}$ is the geminate recombination yield
for the equilibrium initial $\rho_g = \rho_{_E}$, $\,\hat
{\,\mathbf{P}}_{\!f}^{}$ is the superoperator of reaction/relaxation
probabilities defined in eq. (\ref{mfe23}), and $K_{f}^{0} = 4\pi
Dd$ is the rate of PP contacts.

Note that the dimensionless rate $K_{{f}}^{}/K_{f}^{0}$ can be
related to the yield $Y_{{f}_T^{}}$ of the geminate PP recombination
for (triplet) initial density matrix $\rho_g = \rho_{_T}$:
\begin{equation} \label{bulk7}
K_{{f}}^{}/K_{f}^{0} = Y_{f_E^{}}^{} = \mbox{$\frac{1}{4}$}{\cal
P}_{r_{f}^{}}^{_S}\big(1 + 3 Y_{{f_T^{}}}\big),
\end{equation}
where ${\cal P}_{r_{f}}^{_S} = l_{SS}^{}/d $ is the PP recombination
probability in $S$ state. Equation (\ref{bulk7}) is obtained using
the relation $\hat {\,\mathbf{P}}_{\!f}^{} = \hat {\cal
P}_{r_{\!f}^{}}(1 + d\hat k_{0}^{}\hat {\cal P}_{e_{\!f}^{}})^{-1} =
\hat {\cal P}_{r_{\!f}^{}} - \hat {\,\mathbf{\,P}}_{\!f}^{}(d\hat
k_{0}^{}\hat {\cal P}_{e_{\!f}^{}}).$

\subsection{Interacting polarons}

In the considered case of attractive interaction the expression for
the PP recombination rate $K_{c}^{}$ can be obtained within the two
state model (Sec. VI.B). For bulk reactions the spin density matrix,
represented in terms of spin density matrices of the state within
the well $n_0^{}$ and the free diffusion state $\sigma_0^{} (r)$,
satisfy the steady state variant of eqs. (\ref{mfe36a}) and
(\ref{mfe36b}), i.e. equations with $\dot n_0^{} = \dot \sigma_0^{}
= 0$, but with the outer boundary condition: $\sigma_0^{}(r \to
\infty) = r \rho_{_E}$.

Solution of these steady state equations leads to the following
expression for the PP recombination rate
\begin{equation} \label{bulk8}
K = K_{c}^{} = K_c^{0}Y_{c_E^{}} \;\;\mbox{with} \;\; Y_{c_E^{}} =
\langle SS| \hat {\,\mathbf{P}}_{\!c}^{} |\rho_{_E}^{}\rangle,
\end{equation}
in which $ Y_{c_E^{}}$ is PP recombination yield for the equilibrium
(spin) initial state, $\hat {\,\mathbf{P}}_{\!c}^{}$ is given by eq.
(\ref{mfe41}), and $K_c^{0} = 4\pi Dl_c$ is the rate of capture into
the well (cage).

The relation between the dimensionless rate $K_{{c}}^{}/K_c^{0}$ and
the recombination yield $Y_{c_{_T}}$ [for the triplet ($T$) initial
condition], similar to eq. (\ref{bulk7}), can also be derived in the
case of attractive interaction:
\begin{equation} \label{bulk9}
K_{{c}}^{}/K_c^{0} = Y_{c_E^{}}^{} = \mbox{$\frac{1}{4}$}{\cal
P}_{r_{c}^{}}^{_S}\big(1 + 3 Y_{{c_T^{}}}\big),
\end{equation}
where ${\cal P}_{r_{c}}^{_S} = \langle SS |(\hat W_r/\hat W_c)|SS
\rangle $ is the probability of in-cage reaction in $S$ state. The
derivation is based on equation $\hat {\,\mathbf{P}}_{\!c}^{} = \hat
{\cal P}_{r_c} - \hat {\,\mathbf{P}}_{\!c}^{}[\hat \Lambda/W_e +
(l_c\hat k_{0}^{})]\hat {\cal P}_{e_c}$, with $\hat {\cal P}_{r_c}$
and $\hat {\cal P}_{e_c}$ defined in eq. (\ref{mfe42a}).

\section{Results and discussion}

\subsection{General remarks}

\subsubsection{Validity of approaches}

Before the analysis of the MFEs it is worth adding some comments on
validity and accuracy of the proposed mechanisms of spin relaxation.
In the applied model the relaxation, assumed to be induced by
fluctuating HFI and AZI, is described by simple Bloch-type
equations, which are valid at times longer than the correlation
times $\tau_h$ and $\tau_a$ of HFI and AZI fluctuations. This means
that the relaxation mechanisms are applicable only if the time of
MFE formation $\tau_{_{S}} \gg \tau_h,\tau_a$.

In the free diffusion model the MFE-formation time $\tau_{_S}$ is
completely determined by the time of spin evolution in polarons:
$\tau_{_S} \sim \| \hat \Lambda\|^{-1} \sim
w_n^{-1},w_p^{-1},(\Delta \omega)^{-1}$.\cite{Shu1} Taking into
account that the relaxation rates ($w_{_H}$ and $w_{_A}$) and
coherent evolution frequencies ($\sim \Delta \omega$)] are typically
of order of (or less than) $10^{8}\, {\rm s^{-1}}$,\cite{St} one
obtains: $\tau_{_S} \gtrsim 10^{-8} \,{\rm s}$.

Similar estimation for $\tau_{_S}$ is valid in the diffusive cage
model. The additional time parameter, which could change it, is the
characteristic inverse rate of the cage decay $\tau_c = \| \hat
W_c^{}\|^{-1}$. However, this time is also fairly long: $\tau_c
\gtrsim w_0^{-1} e^{u_a}$, where $w_0$ is the characteristic hopping
rate and $u_a$ is the activation energy for escaping and/or reaction
processes. The assumption of deep well ($u_a \gg 1$) leads to the
estimation $\tau_c \gg w_0^{-1}$.

As for correlation times they are expected to be of order of inverse
hopping rates: $\tau_h \sim \tau_a \sim w_0^{-1}$, since HFI and AZI
fluctuations result from stochastic polaron hopping. Typically the
rate $w_{0}^{} \sim 10^{9} - 10^{11} \,{\rm s}^{-1}$, which
corresponds to room-temperature mobilities $\mu_{e,h} \sim 10^{-8}-
10^{-5} \,{\rm cm^2/(V\,s)}$,\cite{Woh4} therefore we get
$\tau_{h,a} \sim 10^{-9} - 10^{-11} {\rm s}$.

Comparison of these estimations shows that the above-mentioned
validity criterion, $\tau_{_{S}}, \tau_c  \gg \tau_h,\tau_a$, is
fulfilled for a large number of semiconducting systems considered.

Concluding the discussion note that the validity of the diffusion
approximation implies negligibly small contribution (to the MFEs) of
small times $t \sim \tau_0 = w_0^{-1}$, at which the diffusive
motion is not yet formed. In the considered $f$- and $c$-models,
according to the above relations, their MFE contributions $\delta
Y_{\mu} = Y_{\mu}(B) - Y_{\mu}(0), \, (\mu = f,c),\, $ can be
written as $\delta Y_f \sim (w_{0}^{}\tau_{_S})^{-1/2}$ and $\delta
Y_c \sim \tau_c^{}/\tau_{_S} \sim e^{u_a}/(w_0^{}\tau_{_S})$. As to
the contribution $\delta Y_{sm}^{}$ of small times $t \sim \tau_0$,
it is, evidently, represented by $\delta Y_{st}^{} \sim
(w_0^{}\tau_{_S})^{-1}$, i.e. in the considered limit
$w_0^{}\tau_{_S} \ll 1$ the contribution $\delta Y_{st}^{}$ is
really small: $\delta Y_{st}^{} \ll \delta Y_{f}^{}, \delta
Y_{c}^{}$ and the diffusion approximation is applicable.

\subsubsection{Parameters and observables}

In the proposed models $B$-dependence of the recombination yield $Y
(B)$ is determined by a very large number of parameters of the
model. To reduce this number in our illustrative discussion of most
important properties of $Y(B)$, we consider the particular (but
representative) variant of the model, in which population and phase
relaxation rates ($w^p$ and $w^n$, respectively) are the same in
$e$- and $h$-polarons:
\begin{equation}\label{dis1a}
w_{_{Q_e}}^p = w_{_{Q_h}}^p = \mbox{$\frac{1}{2}$}w_{_{Q}}^p; \;\;
\; w_{_{Q_e}}^n = w_{_{Q_h}}^n = \mbox{$\frac{1}{2}$}w_{_{Q}}^n,
\end{equation}
with $Q = H, A$; and
\begin{equation}\label{dis1}
w_{_{Q}}^p = \bar w_{_{Q}}^{} {\cal J}_{_Q}^{}(\omega); \;\ \;\;
w_{_{Q}}^n = p_{_Q}\bar w_{_{Q}}  + w_{_{Q}}^p.
\end{equation}
In formulas (\ref{dis1a}) and (\ref{dis1}) the parameter $Q$ denotes
the contributions of HFI ($Q = H$) and AZI ($Q = A$) mechanisms, and
the parameters $p_{_Q}$ and $\bar w_{_{Q}} $ are defined in eqs.
(\ref{ham7}) and (\ref{ham12}). In these formulas we neglect the
difference of Zeeman frequencies in functions ${\cal
J}_{_Q}^{}(\omega_{\nu}^{})$, i.e. took $\omega_e \approx \omega_h =
\omega = \bar g \beta B$.

The MFEs on geminate and bulk PP recombination are found to be
closely related [see eqs. (\ref{bulk6})-(\ref{bulk9})], so that it
is sufficient to analyze, for example, the MFE on bulk process, i.e.
$B$-dependent (dimensionless) rate $K_{\mu}^{}(B)/K_{\mu}^{0} =
Y_{{\mu}_{_E}}(B)$. In what follows we will discuss the function
\begin{equation}\label{dis1b}
y_{\mu}^{}(B) = [Y_{{\mu}_{_E}}(B)-
Y_{{\mu}_{_E}}(0)]/Y_{{\mu}_{_E}}(0), \;\; (\mu = f, c).
\end{equation}

In our work we have obtained fairly simple matrix expressions for
the MFE $y_{\mu}^{}(B)$ in two models of relative polaron motion.
For qualitative understanding of the properties of these functions,
however, it is of certain interest to get simple approximate
analytical expressions. Below we will derived them in some limiting
models.

\subsection{Simple limiting models}

\subsubsection{Fast freely diffusing polarons.}

In the case of freely diffusing polarons of special interest is the
limit of fast polaron diffusion or slow polaron spin evolution, in
which $\| \hat k_0\|d \ll 1$.

General consideration of the problem predicts for the recombination
yield the expression (\ref{mfe28}). In what follows we will mainly
discuss the shape of the MFE $Y_{{f}_{_E}}$, which can be written as
\begin{equation}\label{dis2}
Y_{{f}_{_E}} \approx \mbox{$\frac{1}{4}$}{\cal P}_{r_f^{}}^{_S} +
\mbox{$\frac{1}{16}$}({\cal P}_{r_f^{}}^{_S})^2 (d k_w),
\end{equation}
where $k_w = [\sqrt{w_{p}^{}}+2{\rm Re}(\sqrt{w_{n}^{} + i
\Delta\omega})]/\sqrt{D}$ [see eq. (\ref{mfe29})], ${\cal
P}_{r_f^{}}^{_S} = l_{SS}/d$ is recombination probability in $S$
state, $w_p$ and $w_n$ are population and phase relaxation rates
defined in eqs. (\ref{mfe31}) and (\ref{mfe32}), respectively, and
$\Delta\omega = (\bar g_e - \bar g_h)\beta B = (\Delta \bar g) \beta
B$ is the difference of Zeeman frequencies [eq. (\ref{mfe30})].
According to formulas (\ref{mfe31}) and (\ref{mfe32}) the rates
$w_p$ and $w_n$ are the sums of the corresponding rates for $e$ and
$h$ polarons, which are, in turn, the sums of contributions of HFI
and AZI mechanisms: $w_p^{} = 2(w_{_H}^p + w_{_{A}}^p)\,$ and $\,
w_n^{} = w_{_H}^n + w_{_A}^n\, $ so that
\begin{eqnarray}%\label{dis4}
w_p^{} (B) &=&  2[\bar w_{_H}^{} {\cal J}_{_H}^{}(B) + \bar
w_{\!_A}^{}\!(B)
{\cal J}_{_A}^{}\!(B)],\label{dis4a}\\
w_n^{}(B) &=& \bar w_{_H}^{} + \mbox{$\frac{4}{3}$} \bar
w_{\!_A}^{}\!(B) + \mbox{$\frac{1}{2}$}w_p^{}(B). \quad\quad\;\;
\label{dis4b}
\end{eqnarray}
In these equations we have introduced the function $\bar
w_{\!_A}^{}\!(B)$ to emphasize that (unlike $\bar w_{_H}^{}$) $\bar
w_{\!_A}^{}$ depends on $B$ [see eq. (\ref{ham17})]: $\bar
w_{_A}^{}(B) \sim B^2_{}$.

Noteworthy is that, although formula (\ref{dis2}) is derived in the
fast diffusion limit $d k_w \ll 1$, it appears to be quite accurate
(within $15 - 20\%$) even at $d k_w \sim 1$.

\subsubsection{The limit of weak reactivity in the well.}

Majority of specific features of MFEs in the presence of the well
(diffusive cage) can be analyzed with the limit of weak reactivity,
when $\zeta_c = \|\hat W_r \|/ W_e \ll 1$. In this limit the
expression for the yield $Y_{c_E^{}}$ can be derived by expansion of
the general formulas (\ref{mfe41a}), (\ref{mfe41aa}), and
(\ref{bulk9}) in small $\zeta_c$. In the lowest order we get
\begin{eqnarray}
Y_{c_E^{}} \!&\approx& \!\mbox{$\frac{1}{4}$}{\cal P}_{r_c}^{_S}
\nonumber  \\
&&\!+\mbox{$\frac{1}{16}$}({\cal P}_{r_c}^{_S})^2 \big\{ N(w_{p}^{})
+ 2{\rm Re}\big[N (w_{n}^{} + i \Delta \omega)\big]\big\},\qquad
\label{dis5}
\end{eqnarray}
where ${\cal P}_{r_c}^{_S} = \langle SS| \hat W_r / \hat W_c
|SS\rangle \ll 1$ is the probability of reaction in the well and
\begin{equation}\label{dis6}
N(\epsilon) = 1-\big[1 + (\epsilon/W_e^{}) + \xi
\sqrt{\epsilon/W_e^{}}\,\big]^{-1}
\end{equation}
with $\xi = l_e^{}\sqrt{W_e^{}/D}$.

It is seen that in the limit of slow spin evolution,
%$\bar {\epsilon}$ $\sim w_p, |w_{n}^{} + i \Delta \omega|$,
when $w_p, |w_{n}^{} + i \Delta \omega| \ll \xi^2 W_e $, we get $\,
N(\epsilon) \sim \sqrt{\epsilon}$ and formula (\ref{dis5}) reduces
to eq. (\ref{dis2}), derived in the free diffusion model. Of great
interest is also the opposite limit, in which the evolution kinetics
is described by simple first order kinetic equations and
$N(\epsilon) \sim (\epsilon/W_e^{})[1 + (\epsilon/W_e^{})]^{-1}$.
This kinetics shows itself in the typical analytical dependence of
MFEs on relaxation rates and the splitting $\Delta\omega$.

\subsection{MFE for different relaxation mechanisms}

In our analysis (to reduce the number of parameters) we consider the
limit of high reactivity at a contact, $\kappa_{_S}^{d}/D \gg 1$, in
which reaction/relaxation supermatrices $\hat l$ and $\hat W_{r}^{}$
are independent of $\kappa_{_S}^{}$:
\begin{equation}\label{dis6a}
\hat l \approx d {\hat {\bar {\cal P}}}_{_S}^{}, \;\; \hat W_r^{}
\approx w_r^{}{\hat {\bar {\cal P}}}_{_S}^{} \;\;\mbox{with} \;\;
{\hat {\bar {\cal P}}}_{_S}^{} = 2{\hat {{\cal P}}}_{_S}^{} - \hat
P_{_{SS}}^{}.
\end{equation}
Here ${\hat {{\cal P}}}_{_S}^{}$ is defined in eq. (\ref{mfe7b}) and
$w_r^{} = D/(\kappa_r^{} Z_w^{})$.

We also discuss in detail the special case of Lorenzian shape of
functions ${\cal J}_{_{Q}}^{}(B) = (1 + B^2/B_{_Q}^2)^{-1}$, for
which the relaxation rates are conveniently represented as
\begin{equation}\label{dis6b}
w_{_H}^{p}= \bar w_{_H}^{}/(1 + b_{_H}^2) \;\mbox{and}\; w_{_A}^{p}=
\bar w_{_A}^{*}b_{_A}^2/(1 + b_{_A}^2),
\end{equation}
where $b_{_Q} = B/B_{_Q}$ with $B_{_Q}$ being the width of ${\cal
J}_{_{Q}}^{}(B)$.

\subsubsection{HFI induced relaxation}

\begin{figure}
\setlength{\unitlength}{1cm}
\includegraphics[height=11cm,width=8cm]{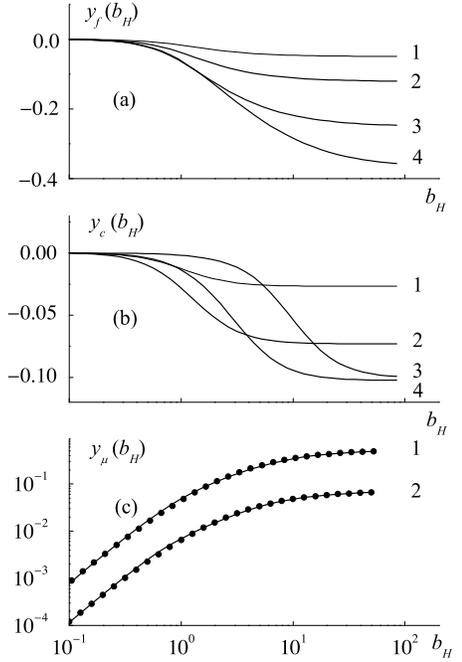}% Here is how to import EPS art
\caption{Dependences $y_{\mu}^{}(b_{H}^{})$ on $b_H^{} =
B/B_{_H}^{}$ for the HFI induced relaxation mechanism, calculated in
free diffusion $(\mu = f)$ and cage $(\mu = c)$ models: for
Lorenzian (Figs 2a and 2b) and non-Lorenzian (Fig. 2c) shapes of
${\cal J}_{_H}^{}(B)$. Fig. 2a shows $y_{f}^{}(b_{H}^{})$ for
different $\eta_{_H}^{} = d \sqrt{\bar w_{_H}^{}/D}$: $\eta_{_H}^{}
= 0.1\: (1); \, 0.3\: (2); \, 1.0\: (3); \, 3.0 \: (4)$. Fig. 2b
shows $y_{c}^{}(b_{H}^{})$ for $r_{c} = w_r^{}/W_c^{} = 0.5$ and
different $\kappa_{_H}^{} = W_c/{\bar w_{_H}}$: $\kappa_{_H}^{} =
5.0\: (1);\, 1.0 \: (2);\, 0.1\: (3);\, 0.01\: (4)$. Fig. 2c shows
arbitrarily scaled $y_{\mu}^{}(b_{H}^{})$ for non-Lorenzian ${\cal
J}_{_H}^{}(B)$ (Sec. VIII.C.1) with $\eta_{_H}^{} = 10^{-4} $ for
$\mu = f \; (1)$ and $r_{_H} = 1.0, \:\kappa_{_H}^{} = 5.0$ for $\mu
= c\; (2)$. Circles represent the prediction of formula $y_{ex}^{}
(B) \sim [B/(B_0^{} + |B|)]^2$ with $B_0=2.35 B_{_H}$, obtained by
fitting $y_{\mu}^{}(b_{H}^{})$.}
\end{figure}

Specific features of dependences $y_{\mu}^{} (B)$ for the
HFI-induced relaxation mechanism are demonstrated in Fig. 2 for the
free diffusion ($\mu = f$) and diffusive cage ($\mu = c$) models. In
both models this mechanism leads to the decreasing functions
$y_{\mu}^{} (B)$. The decrease results, evidently, from that of
population relaxation rates as $B$ increased.

As expected from analytical expressions (\ref{dis2}) and
(\ref{dis5}) these models predict similar dependence $y_{\mu}^{}
(B)$. In a wide region of values of the diffusion coefficient and
reactivity the change of these parameters is found to lead mainly to
the change of the MFE amplitude rather than to that of the shape of
$y_{\mu}^{} (B)$-dependence.

The shape is mostly determined by the mechanism of relaxation, or
more accurately by the form of the correlation function ${\cal
J}_{_H}^{}(B)$. For example, in the case of Lorenzian ${\cal
J}_{_H}^{}(B)$ [see eq. (\ref{dis6b})] the MFE shape is close to
Lorenzian with the width of order of that for ${\cal J}_{_H}^{}(B)$.

In our short discussion we are not going to review all properties of
the shape of $y_{\mu}^{} (B)$. Majority of them can be described
with formulas (\ref{dis2}) and (\ref{dis5}). More thoroughly we will
only analyze the possibility of interpretation of the non-Lorenzian
MFE shape $y_{ex}^{} (B) \sim [B/(B_0^{} + |B|)]^2$, observed in
some experiments.\cite{Woh1,Maj}

Note that for the non-Lorenzian function ${\cal J}_{_H}^{}(B)$ the
MFEs $y_{\mu}^{} (B)$ are non-Lorenzian as well. Just this property
of the HFI induced relaxation mechanism allows one to describe the
observed dependence $y_{ex}^{} (B)$. The reaction yields $y_{f}^{}
(B)$ and $y_{c}^{} (B)$, shown in in Fig.2c, are obtained for ${\cal
J}_{_H}^{\mu}(B) , \; (\mu = f, c)$, represented as sums of
Lorenzian contributions $Lo_n^{\mu}(b_{_H}) = [1 +
b_{_H}^2/(j_n^{\mu})^2]^{-1}$, in which $j_n$ are the numerical
coefficients: ${\cal J}_{_H}^{}(B) \sim \sum\nolimits_{n=1}^6
a_n^{\mu} Lo_n^{\mu}(b_{_H}), \; (\mu = f, c)$, where $a_i^{\mu}$
are the weights of the Lorenzian contributions, taken the same both
for the free diffusion and diffusive cage models: $a_i^{f} = a_i^{c}
= 1, \: (i = 1,\dots,6)$. The values of $j_{i}^{f}$ and $j_{i}^{c}$,
conveniently represented as vectors ${\bf j}^{\mu} =
(j_1^{\mu},\dots,j_n^{\mu})$, are different in these models: ${\bf
j}^{f} = (1,2,4,4,6,10)$ and ${\bf j}^{c} = (1,3,4,5,7.5,20)$. The
chosen number of terms, $n = 6$, seems to be the smallest of those,
which are sufficient to get the yields quite close to $y_{ex}^{}
(B)$.

Thus calculated $y_{f}^{} (B)$ and $y_{c}^{} (B)$ agree with
$y_{ex}^{}(B)$ quite well in the wide region of $B$ values.

The obtained wide spectrum of widths, required in both models, is
qualitatively consistent with the wide distribution of hopping rates
(quite natural for disordered semiconductors), which determines the
behavior of correlation functions, according to eq. (\ref{ham0f}).

\subsubsection{AZI induced relaxation}

Unlike the HFI induced relaxation the AZI induced one results in the
increasing dependences $y_{\mu}^{} (B)$ in both models of PP
migration. As for the characteristic features of the shape of these
dependences, they are not very similar to those for HFI induced
relaxation. Typical dependences $y_{f}^{} (B)$ and $y_{c}^{} (B)$
are displayed in Figs. 3a and 3b for Lorenzian shape of ${\cal
J}_{_A}^{}(B)$ [see eq. (\ref{dis6b})].

It is seen from Figs. 3a and 3b that for AZI mechanism the widths of
$y_{f}^{} (B)$ and $y_{c}^{} (B)$ are larger than those for HFI
mechanism and, in general, are independent of the width $B_{_{A}}$
of ${\cal J}_{_A}^{}(B)$. The reason of this independence consists
in that, unlike the HFI induced relaxation, the AZI induced one
results in dephasing, whose rate rapidly increases with $B$:
$w_{_A}^{n} \sim \bar w_{_A} \sim B^2$ [see eq. (\ref{dis1})]. In
such a case the width of $y_{\mu}^{} (B)$ is mainly determined by
the kinetic saturation of the MFEs observed for large dephasing
rates $w_{_{A_{e,h}}}^n$  [Sec. IV.A2 and eq. (\ref{dis1a})]. In the
free diffusion model the saturation occurs at $|k_w^{}|d > 1$, i.e.
outside the region of validity of the fast diffusion limit
(\ref{dis2}) (Sec. VI.A2). As to the cage model, the saturation in
this model is observed for $\|\hat \Lambda \|/\|\hat W_c^{}\| > 1$.
Within the weak reactivity limit (\ref{dis5}) this condition reduces
to simple inequality $w_n^{}/W_e^{} > 1$.

\begin{figure}
\setlength{\unitlength}{1cm}
\includegraphics[height=8.5cm,width=7cm]{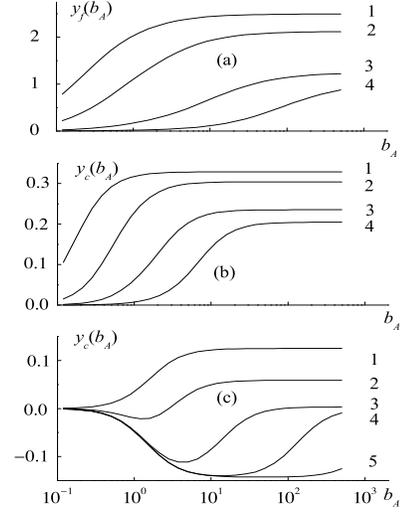}% Here is how to import EPS art
\caption{The dependences $y_{\mu}^{}(b_{A}^{})$ on $b_A^{} =
B/B_{_A}^{}$ for the AZI induced relaxation mechanism and for the
Lorenzian shape of ${\cal J}_{_A}^{}(B)$, calculated in free
diffusion $(\mu = f)$ and cage $(\mu = c)$ models (Figs. 3a and 3b),
and $y_{c}^{}(b_{AH}^{})$  for the HFI+AZI relaxation mechanism
[assuming ${\cal J}_{_H}^{}(B)={\cal J}_{_A}^{}(B)$ and $b_{A}^{} =
b_{H}^{}$] (Fig. 3c). Fig. 3a displays $y_{f}^{}(b_{A}^{})$ for
different $\eta_{_A}^{} = d \sqrt{\bar w_{_A}^{*}/D}$: $\eta_{_A}^{}
= 0.02\: (1); \, 0.2\: (2); \, 2.0\: (3); \, 10.0 \: (4)$. Fig. 3b
displays $y_{c}^{}(b_{H}^{})$ for $r_{c}^{} = w_r^{}/W_c^{} = 0.5$
and different $\kappa_{_A}^{} = W_c/{\bar w_{_A}}^{*}$:
$\kappa_{_A}^{} = 0.05\; (1);\, 0.5 \: (2);\, 5\: (3);\, 50\: (4)$.
Fig. 3c displays $y_{c}^{}(b_{A}^{})$ for $r_{c}^{}  = 1, \:$
$\kappa_{_H}^{} = W_c/{\bar w_{_H}}^{} = 1$, and different
$\kappa_{_A}^{} = W_c/{\bar w_{_A}}^{*}$: $\kappa_{_A}^{} = 1\: (1);
\, 2.25\: (2); \, 4 \: (3);\, 10^2\: (4);\, 10^4\: (5)$.}
\end{figure}

Most clearly the peculiarities of MFEs, resulting from AZI induced
relaxation, manifest themselves in the limits of fast free diffusion
and fast escaping from the cage ($w_{p,n}^{}/W_e^{} \ll 1$). In
these limits obtained formulas predict [see eqs. (\ref{dis2}) and
(\ref{dis5})] monotonically increasing behavior: $y_{f}^{} (B) \sim
|B|$ and $y_{c}^{} (B) \sim B^2$ in the region of $B$ wider than
that ($B \lesssim B_{_A}$) expected from eq. (\ref{dis6b}).

It is worth noting that the dependences close to $y (B) \sim |B|$
has been recently observed in a number of
experiments.\cite{Gi1,Var3,Maj} The proposed analysis allow for
quite reasonable interpretation of this, at first sight, strange
behavior.

\subsubsection{Superposition of HFI and AZI induced relaxation}

In the case of comparable contribution of HFI and AZI induced
relaxation the free diffusion and diffusive cage models predict a
large variety of MFE dependences $y_{\mu}^{} (B), \;(\mu = f, c)$.
In our brief analysis it is hardly possible to describe all types
$y_{\mu}^{} (B)$ behavior. In general, the superposition of HFI and
AZI (HFI+AZI) induced relaxation mechanisms results in non-monotonic
functions $y_{\mu}^{} (B)$. Some representative examples of them, as
applied to $y_{c}^{} (B)$, are displayed in Fig. 3c. Analysis shows
that the resulting shape of $y_{\mu}^{} (B)$ dependence is
essentially determined by the relative widths and relative weights
(i.e. $\kappa_{_H}^{-1} = {\bar w_{_H}}^{}/W_c$ and
$\kappa_{_A}^{-1} = {\bar w_{_A}}^{*}/W_c$) of both mechanisms.

Naturally, as the weight of the AZI contribution increases
$y_{\mu}^{} (B)$ behavior changes from decreasing to increasing.
Note that the change of the sign of the dependence $y_{\mu}^{} (B)$
has recently observed in a number of experiments. The proposed
theory can, in principle, be useful in interpretation of some of
these observations.

In addition to this short discussion few comments are worth-wile,
nevertheless, on the peculiar non-monotonic behavior of $y_{}^{}
(B)$ functions at small $B$, sometime with the small maximum at $B =
0$, which is found for some sets of parameters of models (see the
curve 2 in Fig. 3c). The fact is that such a behavior has recently
been observed in some semiconductor devices.\cite{Var4} In
principle, the authors of this work assume that these non-monotonic
dependences result from the exchange interaction between polarons at
short distances. Our results show that some other mechanisms, for
example the superposition of HFI and AZI induced relaxation, can
lead to a similar $y_{}^{} (B)$ behavior.

\subsubsection{$\Delta g$-mechanism}

The conventional $\Delta g$-mechanism implies the MFE generation
caused by quantum transitions between $S$ and $T$ states of the PP
(\ref{ham19a})- (\ref{ham19c}), which result from the spitting
$\Delta \omega = (\Delta g)\beta B$ of Zeeman frequencies (see Sec.
IV.B).\cite{St} The difference of $g$-factors of $e$- and
$h$-polarons results from the spin-orbital coupling of electron
spins of polarons in semiconductors,\cite{Car1} i.e. the $\Delta
g$-mechanism can be considered as one of manifestations of the this
coupling in MFEs.

In our analysis of $\Delta g$-mechanism we neglect the effect of
other (relaxation) mechanisms discussed above.

The manifestation of $\Delta g$-mechanism in many types of MFEs is
analyzed in detail in a large number of works as applied to
different chemical and physical processes.\cite{St} The contribution
of $\Delta g$-mechanism to the PP recombination yield
$y_{\mu}^{}(B)$ is also discussed in literature. Just this mechanism
is believed to be responsible for $y_{\mu}^{}(B) \sim \sqrt{B}$
behavior found in some semiconductors  at large fields
$B$.\cite{Var3,Maj} Moreover, in a number of works this behavior is
used as a definition of the $\Delta g$-mechanism. Such a definition
is somewhat misleading.

The fact is that $\Delta g$-mechanism, originally describing the
effect of the term of the Hamiltonian $H$ (\ref{ham17}) proportional
to $\Delta \bar g$, predicts different $y_{\mu}^{}(B)$ dependences,
depending on the kinetics of relative motion of polarons:

\paragraph{Free diffusion model.}

In the case of free diffusion the dependence $y_{f}^{}(B)$ can
approximately be represented as\cite{Shu2,Shu4a} $y_{f}^{}(B) \sim
\sqrt{\Delta \omega} (1 + \xi_f^{} \sqrt{\Delta \omega})_{}^{-1},$
where $\Delta \omega = (\Delta g)\beta B$ and $\xi_{f}^{} \sim
d/\sqrt{D}$. This formula shows for $d \sqrt{\Delta \omega/D} \ll 1$
we get the dependence $y_{f}^{}(B) \sim \sqrt{\Delta \omega} \sim
\sqrt{B} $ (in some works considered as the manifestation $\Delta
g$-mechanism), whereas in the opposite limit $d \sqrt{\Delta
\omega/D} > 1$ the dependence $y_{f}^{}(B)$ saturates, i.e
$y_{f}^{}(B) \sim {\rm const}$.

\paragraph{Diffusive cage model.}

In the diffusive cage model we consider the case of deep well, when
the effect of recapture of escaping particles is small. In the deep
well limit the term $\sim \hat k_0^{} \sim \sqrt{\hat \Lambda}$ in
$\hat {\bf P}_c^{}$ [eq. (\ref{mfe41aa})] can be neglected so that
$y_{c}^{} (B)$ dependence (which becomes analytical) can
approximated by\cite{Shu4,Shu5} $y_{c}^{} (B) \sim (\Delta
\omega)^2/ [1 + \xi_c^{}(\Delta \omega)^2]$, where $\xi_c \sim
\|\hat W_c^{}\|_{}^{-2}$ [in the weak reactivity limit $\xi_c \sim
W_e^{-2}$, as it is seen from eq. (\ref{dis6})]. This means that in
the cage model $\Delta g$-mechanism predicts rapidly increasing
behavior of $y_{c}^{} (B)$ at small $B < \|\hat W_c^{}\|/(\bar g
\beta)$: $y_{c}^{}(B) \sim B_{}^2$, saturating at large  $B > \|\hat
W_c^{}\|/(\bar g \beta)$: $y_{c}^{} (B) \sim {\rm const}$.

\medskip

The simultaneous contributions of $\Delta g$-mechanism as well as
HFI- and ASI-induced relaxation mechanisms can result in the
additional specific features of the MFE shape $y_{\mu}^{} (B)$. We
are not going to discuss them in this work, but only mention that
typically $\Delta g$-mechanism strongly contributes at large $B$, at
which contributions of both other mechanisms are nearly independent
of $B$. In this case the change of MFE shape caused by $\Delta
g$-mechanism can easily be identified and described if needed.

\section{Concluding remarks}

In this work PP recombination mechanisms of MFEs in disordered
semiconductors are analyzed in detail. The magnetic field dependent
PP recombination yield is discussed as the most well known example
of the MFE observable. The hopping migration of polarons is assumed
to result not only in the spatial evolution of polarons, but also in
fluctuations of the HFI and AZI of polarons, thus leading to HFI and
AZI induced spin relaxation. In our work we have considered the
manifestation of these two mechanism of MFEs as well as $\Delta
g$-mechanism.

Simple analytical formulas are derived for the recombination yield
$Y (B)$ in two models of polaron migration: the free diffusion model
and the model of diffusion in the well, or diffusive cage model.
Analysis demonstrates that specific features of polaron migration
shows itself in the shape of $Y (B)$. Most important typical
properties of $Y (B)$-dependence for all three MFE mechanisms and in
both models of migration are discussed in detail.

Concluding this discussion we would like to point out two important
points:

1) Above-obtained results show that the width $B_r^{}$ of the
relaxation induced MFE (PP recombination yield) $y_{\mu}^{}(B)$ is
determined by the hopping rate: $B_r^{} \sim w_0^{}/(g\beta),$ i.e.
is much larger than typical HFI: $B_r^{} \gg \sqrt{\langle
{B}_{\nu}^2 \rangle}$.

2) The proposed theory is applied to analysis of the magnetic field
dependent yield of recombination of $e\!-\!h$ PPs, i.e. polarons
with charges of opposite sign. It is clear, however, that obtained
results are quite applicable to description of recombination of
$e\!-\!e$ PPs or $h\!-\!h$ PPs as well. Just this kind of processes
is recently considered as a mechanism of MFEs in organic
semiconductors.\cite{Bob2}

3) This work concerns the analysis of recombination of polarons,
i.e. particles with the electron spin $1/2$. The obtained matrix
formulas are, however, quite general and can be applied to
describing similar processes with participation of particles with
higher spins, for example, triplet exciton-polaron
quenching,\cite{Sw,St,Gi1} triplet-triplet annihilation,\cite{Sw,St}
etc. The analysis of these processes is a subject of future works.

\textbf{Acknowledgements.}\, The work was partially supported by the
Russian Foundation for Basic Research.

%\newpage

%\newpage

\end{document}